\documentclass[]{aa}
\usepackage{graphicx}
\usepackage{epsfig}
\usepackage{longtable}
\usepackage{lscape}

\begin{document}

\title{High resolution imaging of the early-type galaxy NGC\,1380: an insight into the nature of extended extragalactic star clusters \thanks{Table. 4 is only available in
electronic form at the CDS via anonymous ftp to cdsarc.u-strasbg.fr 
(130.79.128.5)
or via http://cdsweb.u-strasbg.fr/cgi-bin/qcat?J/A+A/}}

\author{A. L. Chies-Santos, B. X. Santiago and M. G. Pastoriza}

\offprints{ana.leonor@ufrgs.br} 

\institute{Departamento de
Astronomia, Instituto de F\'{\i}sica, UFRGS. Av. Bento Gon\c
calves 9500, Porto Alegre, RS, Brazil}

\date{Received --; accepted --}

\abstract
{NGC\,1380 is a lenticular galaxy located near the centre of the Fornax Cluster, northeast of NGC\,1399. The globular cluster system of this galaxy was previously studied only from the ground. Recent studies of similar early-type galaxies, specially lenticular ones, reveal the existence of star clusters that apparently break up the traditional open/globular cluster dichotomy.}{With higher quality photometry from HST/WFPC2 we study the star clusters in NGC\, 1380, measuring their magnitudes, colours, sizes and projected distances from the centre of the galaxy.}{We used deep archival HST/WFPC2 in the B and V bands. We built colour magnitude diagrams from which we selected a sample of cluster candidates. We also analysed their colour distribution and measured their sizes. Based on their location in the luminosity-size diagram we estimated probabilities of them being typical globular clusters as those found in the Galaxy.} {A total of about 570 cluster candidates were found down to $V=26.5$. We measured sizes for approximately 200 of them. The observed colour distribution has three apparent peaks. Likewise for the size distribution.  We identified the smaller population as being mainly typical globular clusters, while the more extended objects have small probabilities of being such objects. Different correlations between absolute magnitudes, sizes, colours and location were inferred for these cluster sub-populations.}{Most extended clusters ($R_{eff} > 4$ pc) share similar properties to the diffuse star clusters reported to inhabit luminous early-type galaxies in the Virgo galaxy cluster such as being of low surface brightness and fainter than $M_V \simeq -8$. We also report on a small group of ($R_{eff} \simeq 10 $ pc), $-8< M_V < -6$, red clusters located near the centre of NGC\,1380, which may be interpreted as faint fuzzies.}

\keywords{Galaxies(individual): NGC\,1380 -
Galaxies: star clusters}

\titlerunning{NGC\,1380: star clusters.}

\authorrunning{Chies-Santos et al.}

\maketitle

\section{Introduction}
The study of extragalactic star clusters has proven to be one of the key tools for the understanding of the formation and evolution of galaxies. By studying star cluster systems one can constrain the star formation history and the way the galaxies we observe today formed, merged or accreted smaller parts.

NGC 1380 is an S0 galaxy in the Fornax cluster ((m-M)=31.4 eg. Ferrarese et al. \cite{fer00}), whose globular cluster system (GCS) has only been studied from the ground (Kissler-Patig et al. \cite{kp97}). The authors find two old populations of globular clusters (GCs): a blue population ((B-V)=0.65) similar to the halo GCs of the Milky Way, which is spherically distributed around the galaxy; and a red population ((B-V)=0.94) that follows the stellar light in ellipticity and position angle associated with the bulge and disk of NGC\,1380.

Globular cluster systems of massive galaxies are now well known to have a bimodal colour distribution, indicating two subpopulations of GCs.(eg. Brodie \& Strader \cite{bs06}). These colour differences can be caused by age and/or metalicity differences, due to the well known age-metalicity degeneracy. Recent spectroscopic studies suggest that the colour bimodality is a result of a metalicity difference between two old subpopulations (eg. Strader \textit{et al.} \cite{strader05}). Their existence indicates at least two major star forming events in the history of most massive galaxies.

Faint Fuzzy Clusters (FFs) were discovered in the nearby lenticular galaxies NGC\,1023 and NGC\,3384 (eg. Larsen \& Brodie \cite{lb00}, Brodie \& Larsen \cite{bl02}). They are different from normal globular clusters and open clusters in many aspects. While the effective radius ($R_{eff}$) for globular and open clusters is in the range of 2-3 pc, the $R_{eff}$ of FFs range is 7-15 pc. They tend to have moderately rich metalicities [Fe/H] $ \sim -0.6$, and luminosities in the range of $-7<M_{V}<-5$ and are thought to be old. Furthermore it has been shown that the FFs of NGC\,1023 lie in a fast rotating ring-like structure within the disk (Burkert, Brodie \& Larsen \cite{bbl05}). 

Diffuse Star Clusters (DSCs) have been found in Virgo Cluster early-type galaxies (Peng et al. \cite{peng06}). These objects span a wide range in effective radii $3<R_{eff}<30$ pc, and have low luminosities ($M_{V}>-8$). Their median colours are redder than the red globular cluster subpopulation. They often match the colour of their host galaxy. 

In the present work we study the star cluster system of the S0 galaxy NGC\,1380 with deep archival Hubble Space Telescope Wide Field Planetary Camera 2 (HST/WFPC2) B and V images. We search for star cluster candidates, obtaining magnitudes, colours and measuring sizes.
The paper is organised as follows: In Sect. 2 we describe the observations and the data reduction. In Sect. 3 we present the procedure we followed to obtain the photometry and discuss the colour magnitude diagrams as well as the colour distribution. In Sects. 4. and 5. we describe the procedure applied to obtain sizes and discuss correlations among relevant quantities such as $R_{eff}$, $M_{V}$, (B-V) colour, distance to the centre. In Sect. 6 we present a summary and concluding remarks.

\section{Observations and data reduction}

We used archival B and V images from the HST Proposal 5480,  taken with 
the purpose of studying supernovae SN1992A.
The images were obtained with the Wide Field Planetary Camera 2 (WFPC2) in 
the F439W and F555W filters. There is one pointing and 4 exposures for 
each filter, see table 1.
The spatial scale is $\rm0.046 \arcsec\,pixel^{-1}$ and the field is of
37\arcsec x 37\arcsec~for the planetary camera (PC). For the Wide Field Camera (WFC) the scale is $\rm0.1\arcsec\,pixel^{-1}$, corresponding to a 80\arcsec
x 80\arcsec~field.

\begin{table}
\centering
\renewcommand{\tabcolsep}{1.0mm}
\begin{tabular}{c c c c c}

\hline\hline
Rootname & Filter & Exposure Time (s)\\
\hline
U2BB0701T & F439W & 1200\\
U2BB0702T & F555W & 900\\
U2BB0703T & F439W & 1200\\
U2BB0704T & F555W & 900\\
U2BB0705T & F439W & 1200\\
U2BB0706T & F555W & 900\\
U2BB0707T & F439W & 1200\\
U2BB0708T & F555W & 900\\

\hline
\end{tabular}
\caption{Journal of Observations - HST/WFPC2 images}
\label{jobs}
\centering
\end{table}

\begin{figure}
\resizebox{\hsize}{!}{\includegraphics{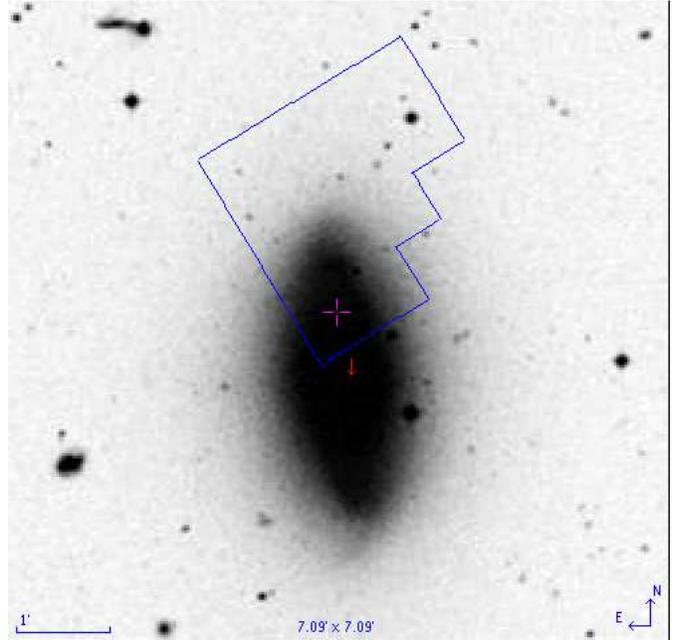}}
\caption[]{A ground based image of NGC\,1380 overploted by the HST/WFPC2 pointing we used in the present study.}
\label{image}
\end{figure}

In Figure \ref{image} we show the WFPC2 field of view which
we were restricted to in the present work.

\section{Photometry}

\begin{figure}
\resizebox{\hsize}{!}{\includegraphics{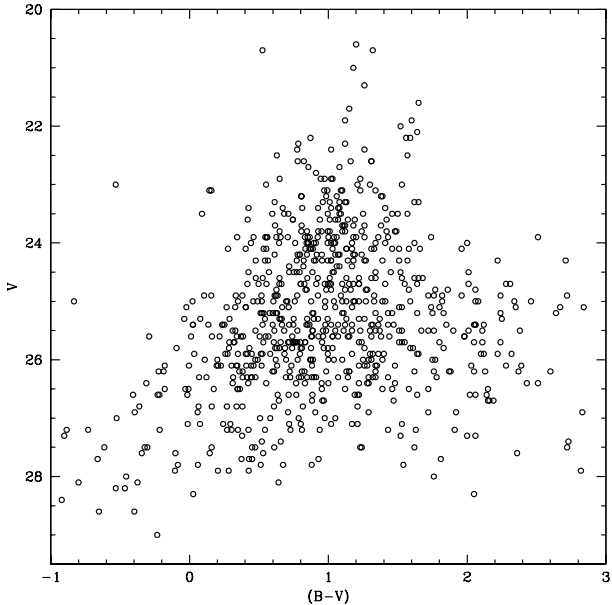}}
\caption[]{Colour magnitude diagram for all sources detected and measured in the image.}
\label{cmd}
\end{figure}

After aligning the 4 individual exposures taken with each
filter, we combined them with the STSDAS task \textit{gcombine}. 
The resulting combined images were almost completely free of cosmic rays. We 
then used the IRAF DAOPHOT task \textit{daofind} to detect sources with a 
threshold of 5 $\sigma$ above the background on the V images. 
Having CCD positions of the detected sources we used the DAOPHOT task \textit{phot} to perform aperture photometry. We adopted a 2 pixel 
aperture radius and applied an aperture correction to a 0.5\arcsec 
radius. The aperture corrections were of 0.31 mag (PC) and 0.18 (WFC) for F439W. For F555W the corrections were of 0.39 mag (PC) and 0.19 mag (WFC). These values were based on Table 2 of Holtzman \textit{et al.} (\cite{holtz95a}). We followed 
Table 7 of Holtzman \textit{et al.} (\cite{holtz95b}) to convert 
the instrumental magnitudes to the standard Johnson-Cousins B,V system, 
for a gain ratio of 7. We then corrected for galactic extinction adopting
the values from Schlegel, Finkbeiner \& Davis (\cite{sfd98}), namely 
$A_{B}=0.075$ and $A_{V}=0.058$. A total of 823 sources were detected
and measured on both images.
In Figure. \ref{cmd} we show the CMD for all these sources. Notice
a strong concentration of sources in the \textbf{$0 < (B-V) < 2$}
range. This locus also spreads out in colour at fainter magnitudes
because of the increasing photometric errors.

\subsection{Colours}

\begin{figure}
%\resizebox{\hsize}{!}{
\vspace{1.0cm}
\includegraphics[scale=0.4]{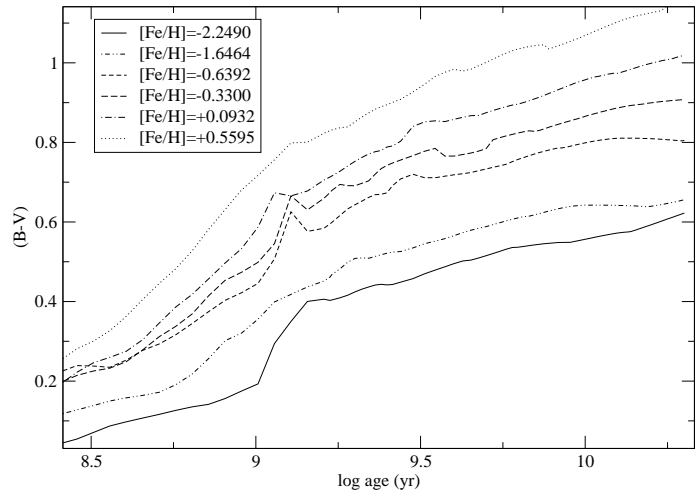}
\caption[]{Evolution of Bruzual \& Charlot (\cite{bc03}) SSPs.}
\label{ssp}
\end{figure}

Simple stellar population models (SSPs), such as Bruzual \& Charlot (\cite{bc03}), can give us an estimate of the expected range in colour of the star 
clusters candidates (Figure. \ref{ssp}).
We assume a wide range in age of $1 - 15$ Gyrs; the lower age limit
assumes that NGC\,1380 has suffered little or no recent
star formation. This is consistent with its red colours and gas paucity. We also consider a wide range in metalicity. The 
corresponding range in colour is $0.2<(B-V)<1.2$. Kissler-Patig et al. \cite{kp97} find that NGC\,1380 GCs have $0.4<(B-V)<1.2$, and that objects 
with $(B-V)>1.3$ are most likely background galaxies. 
However, recent results indicate that the colours of DSCs or FFs may
differ from those of typical GCs. Therefore, a more flexible colour
selection is advisable.
We cut the sample in colour at $-0.1<(B-V)<1.8$ and also apply a cut in magnitude at $V<26.5$, to prevent objects with excessive photometric errors
or spurious sources from contaminating the sample. 
This colour selection criterion accommodates not only the expected
colour range for different SSPs but also the spread caused by photometric
errors and the possible existence of significant internal extinction
in NGC\,1380. With these restrictions we have 570 star cluster candidates. We call these objects as the photometric sample.

\begin{figure}
\resizebox{\hsize}{!}{\includegraphics{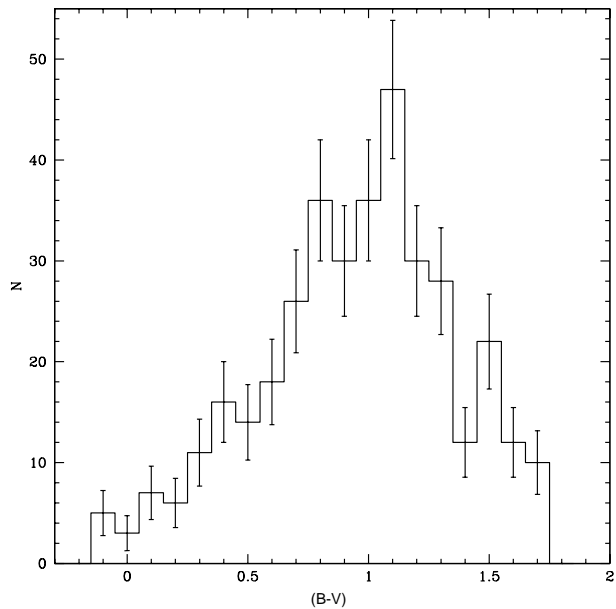}}
\caption[]{Colour distribution of the photometric sample of star cluster candidates.}
\label{bvcol_distribution}
\end{figure}

In Figure \ref{bvcol_distribution} we plot the colour distribution of the photometric sample. Note that there are three peaks at: $(B-V) \simeq 0.8,1.1,1.5$. The first two peaks may be associated to the two different populations of globular clusters found by Kissler-Patig et al. (\cite{kp97}), one metal poor ($(B-V)\sim 0.65$) and the other metal rich ($(B-V) \sim 0.94$). The difference in colour between the peaks shown here and those of Kissler-Patig et al. (\cite{kp97}) is $\Delta(B-V) \sim 0.2$. It can be explained by the clear colour gradient in NGC\,1380 combined with the different regions studied. We will return to this point in Sect. 5. As for the third and very red peak, we also leave its
discussion for latter.

Following Figure \ref{ssp} one can attempt to constrain the ages and 
metalicities that correspond to the distinct cluster sub-populations
according to their (B-V) colours. However, this analysis is
strongly affected by the degeneracy between the effects of varying age and
metalicity. For example, the first colour peak with $(B-V)=0.8$ could be a metal rich population ([Fe/H]=$0.5595$) with $1.3\,Gyr$ 
as well as a sub-solar metalicity population ([Fe/H]=$-0.6392$) 
with $10\,Gyr$. The second and third peaks are definetly due to an old
population with super-solar metalicity, perhaps also affected
by internal extinction.

\section{Sizes}

In order to measure sizes of the star cluster candidates we use the ISHAPE 
code described in Larsen (\cite{larsen99}) and apply it to one of the 
V images listed in Table 2. This code convolves the point spread 
function (PSF) 
with model King profiles in 2 dimensions. The resulting model image 
is then compared to the two-dimensional profiles of the real star 
cluster candidates. We have used the same HST/WFPC2 PSF as in 
Chies-Santos \textit{et al.} (\cite{chies06}). Comparison of this PSF
model to those built from WFPC2 images at other epochs shows that the PSF is
stable with time. In order to derive the best fit 
from ISHAPE we ran the code for King models with 
three different concentration parameter values: c=5, 30 and 100. 
We managed to measure sizes for 302 detected 
sources (out of 823). Applying the previously defined cuts in colour 
and magnitude yields measured sizes for 198 star cluster candidates. As a rule, successful fits were obtained with the three values of
the parameter {\it c}. We then adopted for each source the King model
which minimised $\chi^{2}$. In most cases this model corresponded to c=100.
It is worth stressing that we did not run ISHAPE on the \textit{gcombined} images; instead we ran it on the single exposure U2BB0708T, see Table 1,
even though we loose S/N by doing so.
The reason is that, in the process of image combining, we end up 
degrading the image resolution by broadening the sizes of point sources. 
This would inevitably lead to over-estimates of the clusters sizes with
ISHAPE.

\subsection{Galaxy contamination}

\begin{figure}
\resizebox{\hsize}{!}{\includegraphics{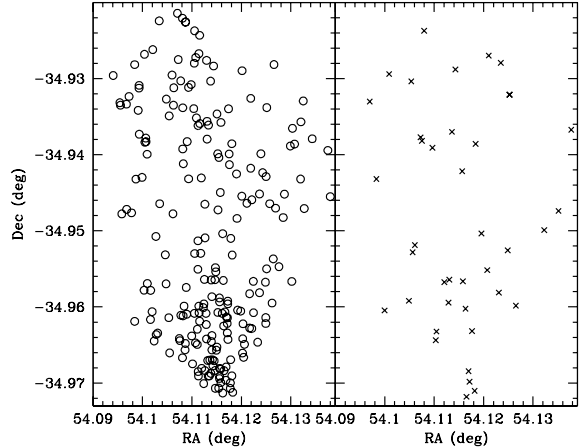}}
\caption[]{Left panel: distribution on the sky of the resolved cluster sample ($R_{eff} < 15$ pc). Right panel: distribution on the sky of the supposedly background contaminants ($R_{eff} > 15$ pc).}
\label{radec}
\end{figure}

Our sample of resolved objects (i.e., for which we could measure sizes) should be contaminated by background galaxies. This effect needs to be 
quantified. The most likely contaminants will be early-type galaxies 
close to the detection limit of our photometry, since these
are more likely to be taken as stellar or nearly stellar objects.
We therefore estimate their numbers by scaling down the number of 
E/S0 galaxies in the morphological sample of Abraham et al. (\cite{abraham96}). These authors provide morphology of galaxies in the Hubble Deep Field 
down to I=25 ($V \simeq 26$). About $4 \times 10^{4} deg^{-2}$ E/S0 galaxies 
are expected within the $23<I<25$ range, which in turn
corresponds roughly to $24<V<26$. Scaling to the WFPC2 solid angle,
this corresponds to about 40 galaxies in our sample. Interestingly, 
this is very similar to the number of sources with $R_{eff}>15$ pc we find, 
assuming that they are located at the distance of NGC\,1380.
Thus, such objects are consistent with being background early-type galaxies
rather than intrinsically very large clusters belonging to NGC\,1380.
We therefore cut the sample at $R_{eff} < 15$ pc in order to 
concentrate on the cluster population. We call this set of objects as the resolved cluster sample.
In Figure. \ref{radec} we show the on-sky distribution of the resolved cluster sample as well as of the objects identified as background galaxies. Notice that while the former objects display a significant gradient towards the centre of the galaxy (towards the bottom) the later do not show this effect so clearly.

Notice that this approach of 
removing background galaxies relies on the assumption that they tend to
be apparently more extended than stellar clusters.
A few very distant early-type galaxies may therefore be still 
left in our star cluster sample.
We also point out that some of the objects with $R_{eff}>15$ pc 
could be dwarf galaxies belonging to the Fornax cluster.

%On the other hand, if they are not, one can analyse the effect of such objects with $R_{eff}>15$ pc not being at the distance of NGC\,1380 by plotting their colours separately.
%One might look at this plot and think that the bimodality has vanished. This is not true though, note that this distribution is only of objects which ishape could resolve, therefore there are still many more star clusters not resolved.

\begin{figure}
\resizebox{\hsize}{!}{\includegraphics{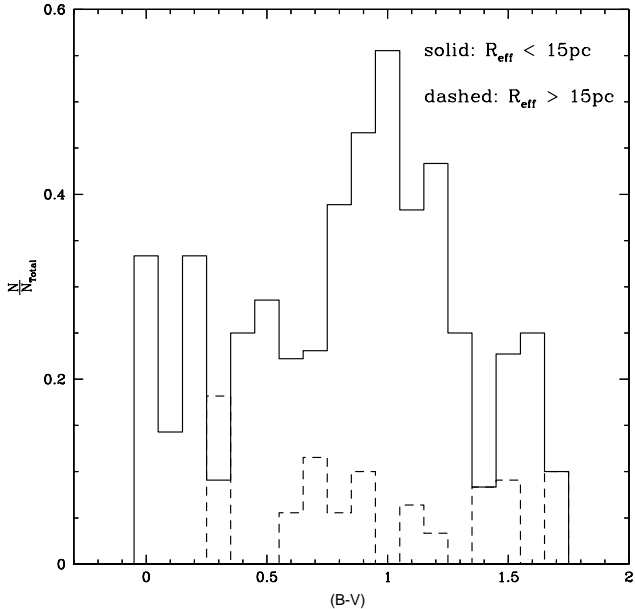}}
\caption[]{Ratio between resolved star clusters and galaxies to the photometric sample as a function of colour. Dashed line corresponds to the objects with $R_{eff}>15$ pc while the solid line to the objects with $R_{eff}<15$ pc.}
\label{bvcol}
\end{figure}

In Figure \ref{bvcol} we plot the number ratio of
the resolved cluster and background galaxies to the photometric sample, as a function of colour.
One can see that this fraction is somewhat smaller for the blue cluster
sub-population, in fact resulting in the elimination of the $B-V \simeq 0.8$
peak seen in Figure. \ref{bvcol_distribution}. In other words, a larger fraction of blue clusters are unresolved and
therefore drop out of the sample with successfully measured sizes.
This result points to a size-colour trend in the sense that redder
GCs tend to be larger than the bluer ones. This issue will be discussed
in more detail latter.
Also notice that objects with $R_{eff}>15$ pc cover a wide
colour range, but have a proportionally larger contribution to the
very red population of cluster candidates in the range $1.35<(B-V)<1.7$.
In the $0.4<(B-V)<1.3$ range they account for $\sim 15\%$ of the sample. 
%Thus, the exclusion 
%of objects with $R_{eff}>15$ pc excludes mainly star cluster candidates 
%bluer than $(B-V)\sim 0.7$.

\subsection{Size and luminosities distributions}

\begin{figure}
\resizebox{\hsize}{!}{\includegraphics{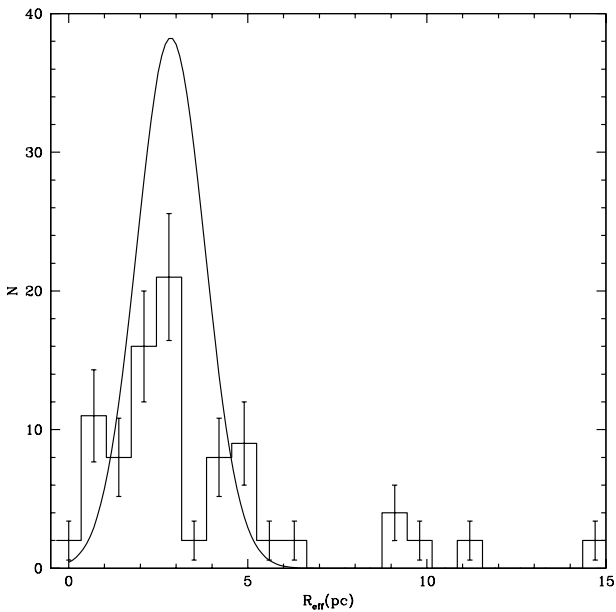}}
\caption[]{The distribution of the resolved star cluster sample ($R_{eff}<15$ pc). Overploted is the size distribution of galactic GCs (Bonatto et al. \cite{bbcps06}).}
\label{reff_distribution}
\end{figure}

In Figure \ref{reff_distribution} we plot the distribution of the
159 resolved clusters (objects with $R_{eff}<15$ pc) belonging to 
NGC\,1380. This distribution clearly shows distinct
peaks. The first peak, at $R_{eff} \sim 3$ pc, closely matches
typical GC sizes, as attested by the solid curve, which corresponds
to the distribution of sizes of GCs in the Galaxy (see below).
The second peak is at $R_{eff} \sim 5$ pc; this
may be the population of DSCs.
There are some objects with $7.5<R_{eff}<13$ pc that could also be the 
diffuse star clusters reported in Peng et al. (2006) or FFs.

It is very important to characterise the possible families
of stellar clusters present in NGC\,1380 based on their magnitudes,
colours, sizes and spatial location. In particular, one of our goals
is to search for possible examples of DSCs or FFs in this galaxy.

Thus, we now compare the available luminosity and size distributions
of our clusters with the known distributions of typical GCs as a function of 
these parameters. We adopt the parametric GC
luminosity and size distributions given by Bonatto et al. (\cite{bbcps06}).
In this work the authors fit gaussian functions both to the
distribution of half-light radii and of absolute magnitudes, $M_{V}$, 
for the whole sample of Galactic globular clusters. They find

$$N=(40.1 \pm 9.2) exp-\frac{1}{2}[\frac{R_{eff}-(2.85 \pm 0.18)}{0.95 \pm 0.14}]^2, \eqno (1) $$

for the first and

$$N= (19.66 \pm 1.77) exp-\frac{1}{2}[\frac{M_{0}+(7.74 \pm 0.1)}{1.35 \pm 0.08}]^2, \eqno (2) $$ for the second.

The size distribution fitted by those authors is overplotted to the
distribution of NGC\,1380 clusters in Figure \ref{reff_distribution}.
The curve has been re-normalised to the number of star cluster candidates 
of NGC\, 1380. It is clear from the figure that the secondary
peaks at $R_{eff} = 5$ pc and $7 < R_{eff} < 13$ pc are not well 
accounted for by the distribution of Galactic GCs. The $R_{eff} = 5$ pc peak is at the tail of the gaussian 
and the later peak is well away from it.

\begin{figure}
\resizebox{\hsize}{!}{\includegraphics{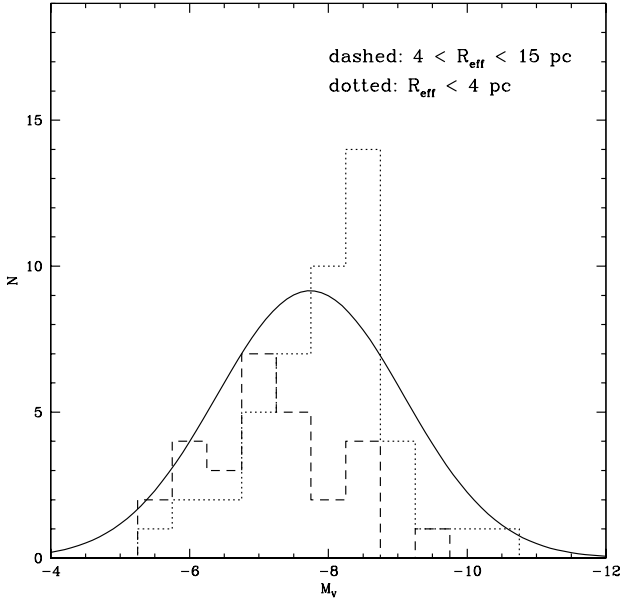}}
\caption[]{Luminosity distribution for the resolved cluster sample
for objects with $R_{eff}<4$ pc and $4 < R_{eff} < 15$ pc. Overploted is the distribution of galactic GCs (Bonatto et al. \cite{bbcps06}) normalised to the dotted histogram.}
\label{mv_distribution}
\end{figure}

In Figure \ref{mv_distribution} we plot the luminosity distribution of the resolved
star cluster sample separated by different intervals in $R_{eff}$. Overplotted is the 
luminosity distribution of the galactic globular clusters 
(Bonatto et al. \cite{bbcps06}) normalised to the number of star clusters
of NGC\, 1380 with $R_{eff} < 4$ pc. It is clear from the figure that the more extended clusters tend
to have lower luminosities than those with typical GC sizes ($R_{eff} < 4$ pc). These latter, which make
up the main peak in the size distribution,
are likely to be the genuine sample of GCs in this galaxy. Notice that neither distributions shown in the figure
are perfectly described by the fit from Bonatto et al. \cite{bbcps06}. The GC-like clusters display a more pronounced peak and smaller dispersion relative to the Galactic counterparts.
There is also a lack of luminous clusters in these distributions relative to what is seen in the Galaxy. This may be a selection effect due to mass segregation; i.e. the largest, and thus most luminous, clusters should be located towards the centre of NGC\,1380 and are therefore not sampled in the image.

We used the available luminosity and size information for our clusters,
coupled with the known distribution of GCs as a function of these parameters,
in order to estimate the probability of a star cluster candidate 
being a globular cluster. We adopted the parametric GC
luminosity and size distributions given by Bonatto et al. (\cite{bbcps06}).

Thus, under the assumption that the GCs in NGC\,1380 follow the same
intrinsic distributions, it is then easy to compute a relative probability
of each cluster being a GC, $P_{GC}$, whose value will be $P_{GC}=1$
for clusters with $R_{eff} = \overline{R_{eff,MW}}=2.85$ pc and 
$M_V = \overline{M_{V,MW}}=-7.74$.

\begin{figure}
\resizebox{\hsize}{!}{\includegraphics{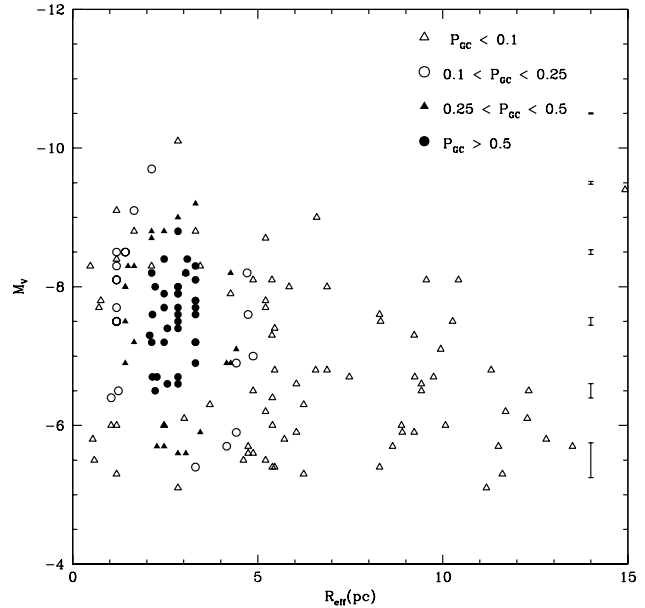}}
\caption[]{$R_{eff}$ as a function of $M_{V}$. Different symbols indicate different GC probabilities, as indicated. The error bars indicate mean errors in $M_{V}$.}
\label{mvreff}
\end{figure}

In Figure \ref{mvreff} we plot the $R_{eff}$ as a function of $M_{V}$.
The 3 distinct peaks seen in the size distribution are
also clearly seen here, with noticeable gaps at $R_{eff} \simeq 4$pc
and $R_{eff} \simeq 7-8$pc. For small radii, $R_{eff} < 3$pc,
ISHAPE finds somewhat discretized solutions for the sizes.
The different symbols indicate different values of $P_{GC}$.
Objects with the greatest probabilities of being a GC, $P_{GC} > 0.5$, 
are located at $R_{eff} \sim 2.5$ pc and $M_{V}\sim-7.5$, as expected.
Even though there is no strong correlation between size and luminosity,
the upper-limit in luminosity varies as a function of $R_{eff}$,
in the sense that the more extended clusters, with low $P_{GC}$ values,
tend to be less luminous.
Note that the lower limit in luminosity ($M_{V} \sim -5$) is due to the cut at 
$V < 26.5$ but the upper limits are physical. 
A slight trend of the $R_{eff} \simeq 5$ clusters concentrating at
lower luminosities than typical GCs is also observed and has
been discussed before.

Previous studies (eg. Larsen et al.\cite{larsen01}, Jord\'an et al. \cite{jordan05}) have found that the GC bluer population 
is $\sim 20\%$ larger than the red population. This fact has been 
interpreted as the result of projection effects, mass-segregation or the 
dependence of stellar lifetimes on metalicity (Larsen \& Brodie \cite{lb03}, 
Jord\'an \cite{jordan04}).
In Figure \ref{colreff} we plot the $R_{eff}$ as a function of (B-V) colour. 
This apparent trend between size and colour is readily seen from the
figure, but seems to be largely due to the contribution of
clusters with $R_{eff} > 5$pc and $P_{GC} < 50\%$.
Considering all the star clusters with $R_{eff} < 15$ pc, the blue 
clusters are in fact $\sim 20\%$ larger than the red ones. When we separate 
the sample in size intervals the trend is reversed: for $0<R_{eff}<4$ pc 
the red GCs are $\sim 5\%$ larger than the blue ones; 
for $4<R_{eff}<8$ pc the red clusters are $\sim 3\%$ larger and 
from $8<R_{eff}<15$ pc the red ones are $\sim 5\%$ larger.\footnote
{The weighted average of the sample split in size bins
is the same as the average of the entire sample.} 
However, these differences are not statistically significant,suggestingg
that the effect of blue clusters being larger than red ones stems
from the fact that they are more common among the extended
objects, with low $P_{GC}$ values.
The detailed out
statistics on the colour-size relation for the NGC\,1380 clusters is
shown in Table 2, where mean sizes and dispersions around this mean
are shown for different colour and size bins.
One has to be very cautious when analysing the observed
differences due to the small numbers often involved.

We also notice the existence of a separate set of red 
clusters ($(B-V) > 1$) with $R_{eff} > 9$ pc, which clearly
detaches from the trend just mentioned. We discuss them in the next section.

\begin{table}
\centering
\renewcommand{\tabcolsep}{1.0mm}
\begin{tabular}{c c c c c c c}

\hline\hline

Blue: $-0.1<(B-V)<0.9$ \\  
& $\overline{R_{eff}}$ (pc)  & $ \sigma_{R_{eff}}$ & NP \\  
   all       &  4.25 &  3.36 & 87\\
$0<R_{eff}<4$ &  1.70 &  1.23 & 45\\
$4<R_{eff}<8$   &  5.23 &  0.83& 27\\
$8<R_{eff}<15$ & 10.15  &  1.79 & 15\\
\\
Red: $0.9<(B-V)<1.8$&\\
& $\overline{R_{eff}}$ (pc)  & $ \sigma_{R_{eff}}$ & NP\\
   all       &3.38 & 3.23 & 99 \\ 
$0<R_{eff}<4$ &1.78 & 1.13 & 73 \\
$4<R_{eff}<8$   &5.43 & 0.82 & 14 \\
$8<R_{eff}<15$ &10.72 & 1.66 & 12 \\

\hline
\end{tabular}
\caption{The sizes of the star cluster candidates separated by colours.}
\label{sizecol}
\centering
\end{table}

%One might then ask at this point if the bimodal colour distribution survives 
%when we consider only typical GCs ($0<R_{eff}<4$ pc). Figure \ref{gcs} shows 
%this distribution. We can see that the difference between the blue peak 
%($(B-V) \simeq 6$) and the red peak ($(B-V) \simeq 1$) is now less pronounced 
%than in Figure \ref{bvcol_distribution}. The larger star clusters are then 
%more metal-poor or younger than typical GCs.

%\begin{figure}
%\resizebox{\hsize}{!}{\includegraphics{gc_bvcol_distribution.eps}}
%\caption[]{Globular cluster colour distribution ($0<R_{eff}<4$ pc).}
%\label{gcs}
%\end{figure}

%Since the exclusion of objects with $R_{eff}>15$ pc excludes mainly star 
%cluster candidates bluer than $(B-V)\sim 0.7$, the fact that we don't find a 
%trend between colour and size for the GCs $R_{eff}<4$ pc can reflect a 
%selection effect associated with a reverse trend that is generally found for 
%other galaxies (REFERENCES!).

\section{Spatial Gradients}

\begin{figure}
\resizebox{\hsize}{!}{\includegraphics{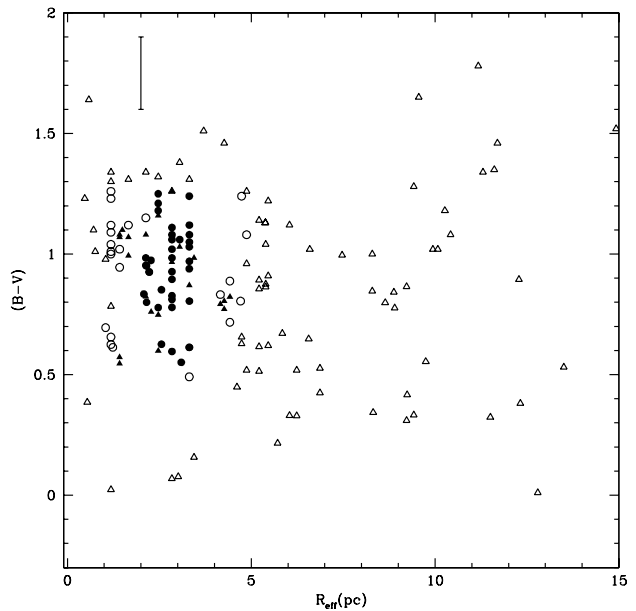}}
\caption[]{$R_{eff}$ as a function of (B-V) colour. Different symbols indicate different GC probabilities as in Figure \ref{mvreff}. The error bar corresponds to a typical colour error.}
\label{colreff}
\end{figure}

\begin{figure}
\resizebox{\hsize}{!}{\includegraphics{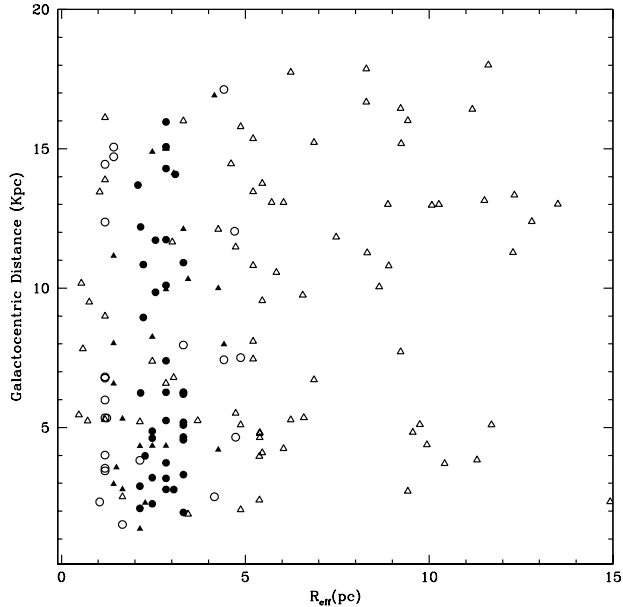}}
\caption[]{$R_{eff}$ as a function of the galactocentric distance. Different symbols indicate different GC probabilities as in Figure \ref{mvreff}.}
\label{reffdistcent}
\end{figure}

\begin{figure}
\resizebox{\hsize}{!}{\includegraphics{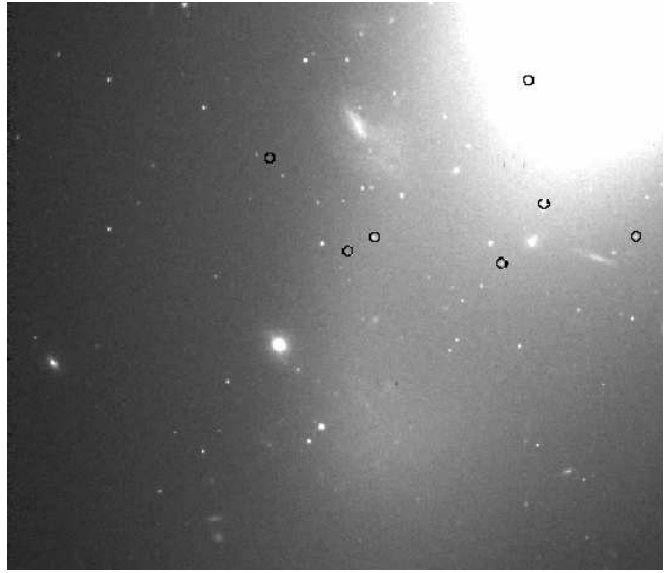}}
\caption[]{HST/WFPC2 ch4. White circles are the 7 FF candidates.}
\label{7dwarfs}
\end{figure}

We now consider possible trends in cluster properties as a function 
of projected galactocentric distance. In Figure \ref{reffdistcent} we plot 
the $R_{eff}$ as a function of this parameter. Objects with 
$P_{GC} > 0.5 $ are slightly more concentrated towards the centrer, 
whereas clusters with the least 
probability of being GCs are spread at the outer parts 
of the galaxy.
This might be caused by the fact that smaller clusters are located preferentially towards the centre, as observed in other galaxies (Larsen \& Brodie \cite{lb03} and Jord\'an et al. \cite{jordan05}).
Note that there is a general trend in the upper limit of the sizes as
a function of galactocentric distance ($<10$ kpc) in the
sense that the upper size limit increases as we go further out in 
galactocentric distance. This is what should be expected due to tidal effects. 
On the other hand there is a group of 7 clusters at the bottom right side 
of the figure that violate the general trend. Interestingly, they
share other common properties: all but one of them are red clusters,
seen at the top right part of Figure \ref{colreff}. 
They are also very near each other in terms of projected distance, 
falling at the same WFPC2 chip. Their positions on the image are shown 
in Figure \ref{7dwarfs}.

\begin{figure}
\resizebox{\hsize}{!}{\includegraphics{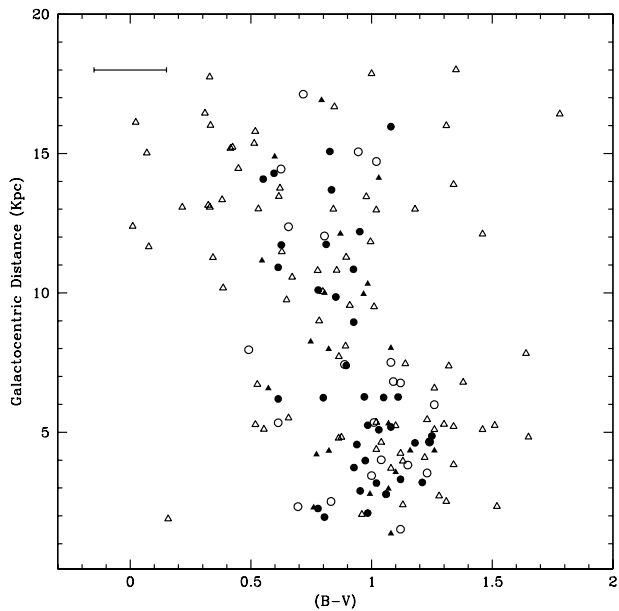}}
\caption[]{Colour as a function of the galactocentric distance. Different symbols indicate different GC probabilities as in Figure \ref{mvreff}. The error bar corresponds to a typical colour error.}
\label{bvcoldistcent}
\end{figure}

Finally, in Figure \ref{bvcoldistcent} we find a trend between colour and 
galactocentric distance in the sense that the bluer star clusters are 
located towards the outer parts of the galaxy. This trend can either reflect 
a metalicity gradient (inner clusters are more metal rich) or the possibility 
of star formation occurring later on a disk than on the spheroidal 
part of the galaxy. The first hypothesis is supported by the fact that many studies 
of early type galaxies indicate an excess of heavier elements (Fe, Mg) in relation to Hydrogen in the centre 
(eg. Rickes, Pastoriza \& Bonatto \cite{rpb05}). Notice that this colour gradient may go a long way towards explaining the colour differences between our colour peaks and those found by Kissler-Patig et al (\cite{kp97}), since these later use ground based images extending much further away from the centre of NGC\,1380, therefore sampling in average bluer objects.

It seems that both the colour and size distributions are
bimodal, possibly trimodal. The colour bimodality is well known 
(e.g Brodie \& Strader \cite{bs06}) and likely reflects different events of star formation
along the history of the host galaxy.
As mentioned earlier, the bimodality in the sizes may be interpreted 
as evidence of the GC and
FF/DSC populations. The later objects tend to be less luminous and more
extended. But the two populations
are not clearly distinct in terms of colours, although the
spread in colours seems larger in the FF/DSC population (low $P_{GC}$) than in 
the GC population (high $P_{GC}$) (Figure \ref{bvcoldistcent}).

%Here we find that the fraction of FFs/DSCs seems to be larger in the blue 
%($(B-V) ~\sim 0.6$)
%peak than in the red (dominating) peak.

\section{Summary and concluding remarks}

We have analysed deep WFPC2/HST images of NGC\,1380 in the B and V filters, from which a sample of about 570 star cluster candidates was drawn. The colour distribution contains three distinct peaks, two of which have been observed in previous works form the ground (Kissler-Patig et al. \cite{kp97}). In order to better characterise these cluster subpopulations we attempted to measure sizes. We also identify three distinct peaks in the resulting size distribution, the main of them being similar to typical globular clusters, whereas the other two corresponding to more extended clusters. We compare both the luminosities and sizes of our sample to the Galactic counterparts: the extended NGC\,1380 clusters are not well fit by the Galactic clusters size distribution; the extended clusters also tend to be less luminous than typical GCs in the Galaxy. We assign a probability of each cluster in our sample to be a GC, $P_{GC}$. We then search for correlations among luminosities, sizes, colours, location and $P_{GC}$. No strong $M_V$ versus $R_{eff}$ correlation was found although the upper luminosity limit varies as a function of size in the sense that the extended sub-populations rarely reach $M_{V} < -8$. As for the size-colour relation we confirm that bluer objects tend to be larger than redder ones. However, this result applies to the entire cluster sample; it ceases to be true when GC-like ($R_{eff}<4$ pc) clusters are considered in separate. No size-colour relation is seen when the extended sub-populations are analysed alone. In fact, we observe that a larger fraction of blue GC-like clusters in the photometric sample drop out when sizes are measured, indicating that among GCs an inverse trend (bluer clusters being smaller) should hold.

We also explore correlations with projected galactocentric distance. We observe that GC-like objects ($P_{GC}>0.5$) are more concentrated towards the central regions. Also, the observed upper limit in size increases with distance from the centre. These results are consistent with previous studies that find that smaller clusters are preferentially located closer to the centre, something that may result from tidal effects.
We also observe a trend of bluer colours with increasing galactocentric distance. This could either reflect a metalicity gradient or an age gradient. The trend is more pronounced for the GC-like objects ($P_{GC}>0.5$) whereas for the other sub-populations there is a larger spread in this correlation.

It is very important to interpret our results in light of the new star cluster populations, such as the FFs and DSCs, that have been recently proposed to exist in the literature. In particular we note that most of the extended clusters ($R_{eff} > 4$ pc) share similar properties as the DSCs found in Virgo early-type galaxies by Peng et al. (\cite{peng06}): they are located at the $\mu_{V} < 20 \, \rm mag\, arcsec^{-2}$ locus in the $M_V$ vs.$R_{eff}$ diagram. They are also upper bounded in luminosity at $M_V \simeq -8$. On the other hand, the candidates found here are not necessarily redder than the redder GCs, as has been previously claimed for DSCs (Peng et al. \cite{peng06}).

We also discovered a subset of $R_{eff} \simeq 10$ pc, $-8 < M_V < -6$, $(B-V) > 1$ clusters located in the inner regions of NGC\,1380 ($2 < R < 5$ Kpc), where R is the projected galactocentric distance). These share similar properties as the FFs found in other lenticular galaxies by other authors. Their main properties are listed in Table. \ref{7dwarfstable}. The whole sample is presented in Table. 4 available in electronic format.

\begin{table}
\centering
\renewcommand{\tabcolsep}{1.0mm}
\begin{tabular}{c c c c c c c}

\hline\hline

RA(deg) & Dec(deg) & V & B & (B-V) & $R_{eff}$ (pc) \\  

54.1171570 & -34.9621773 &  24.325  & 25.328 & 1.020 &  9.94 \\
54.1123238 & -34.9600945 &  25.196  & 26.636 & 1.458 &  11.70\\ 
54.1128502 & -34.9608421 &  23.344  & 24.982 & 1.656 &   9.55\\
54.1206474 & -34.9649086 &  24.643  & 25.966 & 1.340 &  11.30\\
54.1172905 & -34.9642792 &  23.288  & 24.355 & 1.084 &  10.42\\
54.1082268 & -34.9611702 &  24.749  & 25.285 & 0.553 &   9.75\\
54.1145058 & -34.9670792 &  24.809  & 26.070 & 1.277 &  9.42\\

\\

\hline
\end{tabular}
\caption{The subset of extended clusters in Figure \ref{7dwarfs}.}
\label{7dwarfstable}
\centering
\end{table}

\begin{acknowledgements}
We acknowledge the financial support of CNPq. We thank Charles Bonatto and Eduardo Bica for usefull discussions and the anonymous referee for the report.
\end{acknowledgements}

\end{document}